\documentclass[aps,prl,amsmath,amssymb,twocolumn,superscriptaddress,letterpaper]{revtex4}
\usepackage{dcolumn}
\usepackage{graphicx}
\usepackage{epstopdf}
\usepackage{verbatim}
\usepackage{amssymb}
\usepackage{amsmath}
\usepackage{subfigure,wrapfig}
\usepackage{color}

\begin{document}

\title{Air-compatible Broadband Angular Selective Material Systems}

\author{Yichen Shen,$^{1,2\dagger}$ Chia-Wei Hsu,$^{3}$ John D Joannopoulos,$^{1,2}$ and Marin Solja\v{c}i\'{c}$^{1,2}$}
\affiliation{
\normalsize{$^{1}$Department of Physics, Massachusetts Institute of Technology,}\\
\normalsize{Cambridge, MA 02139, USA}\\
\normalsize{$^{2}$Institute For Soldier Nanotechnologies, Massachusetts Institute of Technology,}\\
\normalsize{Cambridge, MA 02139, USA}\\
\normalsize{$^{3}$Department of Applied Physics, Yale University,}\\
\normalsize{New Haven, CT 06520, USA}\\
\normalsize{$^\dagger$To whom correspondence should be addressed; E-mail:  ycshen@mit.edu.}
}

\begin{abstract}

An optical broadband angular selective filter achieved by stacking of one-dimensional photonic crystals has promising potential in various applications. In this report, we demonstrate the first experimental realization of an optical broadband angular selective filter capable of operation in air, in contrast to our previous demonstration in index matching liquid.

\end{abstract}

\maketitle

\section{Introduction}
Light selection (or filtering light) based on the direction of propagation has long been a scientific and engineering challenge \cite{Schwartz:03,Upping2010102,de2007colloquium,le2012broadband,PhysRevLett.106.123902,Atwater_angular2013,Shen28032014,Shen_metamaterial,akozbek2012experimental,argyropoulos2013broadband,argyropoulos2012matching}. Narrowband angular selectivity has been demonstrated in the earlier years using resonance \cite{de2007colloquium}, photonic crystals \cite{Upping2010102}, and metamaterials \cite{Schwartz:03}. Achieving broadband angular selectivity, on the other hand, is more challenging, and has a potential of leading to even bigger fields of applications, such as energy harvesting \cite{Veronika_superlattice_2014,yeng2014global,bermel2011tailoring} and signal detection. Therefore, it has recently started drawing attention from researchers. Methods of using anisotropic photonic crystals \cite{PhysRevA.83.035806} plasmonic Brewster mode\cite{akozbek2012experimental,argyropoulos2013broadband,argyropoulos2012matching} and micro-scale geometrical optics \cite{Atwater_angular2013} have been investigated. However, to the best of our knowledge, yet none of the works have provided an experimental realization in the full visible spectrum. We previously reported broadband angular selectivity of light using photonic crystals (PhC) \cite{Shen28032014} and metamaterials \cite{Shen_metamaterial}. However, the PhC approach requires immersion into index matching gel for proper operation. In this report, we build upon the photonic crystals approach (which can be easily fabricated at large scale), and demonstrate experimentally for the first time a full visible spectrum angular selective filter that is capable of operation in air (i.e. air-compatible) and with adjustable angle of transparency. With improved implementations, we propose two general categories of applications of angular selectivity in the optical regime, with proof-of-concept experimental demonstrations. One application lies in increasing the signal-to-noise ratio (SNR) in light detection systems, while the second application concerns energy harvesting. 

\section{Air-compatible implementation}
Our previous work \cite{Shen28032014} has shown that one can utilize the characteristic Brewster modes in one-dimensional hetero-structured photonics crystals \cite{joannopoulos2011photonic} to achieve broadband angular selectivity. However, the device demonstrated in \cite{Shen28032014} has some limitations. Firstly, it needs to be immersed in an index-matched liquid. In-air operation requires ultra-low index materials as one of the composition layers, which makes it hard to fabricate. Secondly, the Brewster angle at the interface of two dielectric media (in the lower index isotropic material) is always larger than 45$^{\circ}$. Here we introduce an implementation using macroscopic prisms to improve the previous design, such that the device can be directly used in the air without using ultra-low index materials. In addition, the angle of high transparency can be adjusted easily over a broad range, including normal incidence.

Conventional lossless dielectric materials have refractive index $n$ ranging between 1.4 and 2.3 in the visible spectrum\cite{2007PSSBR}. Following the design principle illustrated in Ref.~\citenum{Shen28032014}, if we choose two conventional materials with index $n_1$ and $n_2$ from the above category, assuming $n_1<n_2$, the Brewster angle (defined in material with index $n_1$) is given by:
\[\theta_{B,n_1}=\tan^{-1}\frac{n_2}{n_1}.\]
In order to couple into the Brewster mode in this material system from air, the incident angle (defined in air) needs to be:
\[\theta_{B,air}=\sin^{-1}\left(n_1\sin\theta_{B,n_1}\right).\]
Given that $\theta_{B,n_1}>45^{\circ}$, and $n_1\geq1.4$, the term in the bracket in the right hand side of the above equation is almost always bigger than 1. Therefore, it is challenging to use conventional lossless materials to build an angular selective device that can couple light from air. Porous thin film materials have lower refractive index \cite{2007PSSBR} and can be fabricated with oblique evaporation deposition \cite{so14929}. However, such technology is not yet capable of producing multilayered films.

\begin{figure}[htbp]
\begin{center}
\includegraphics[width=3.5in]{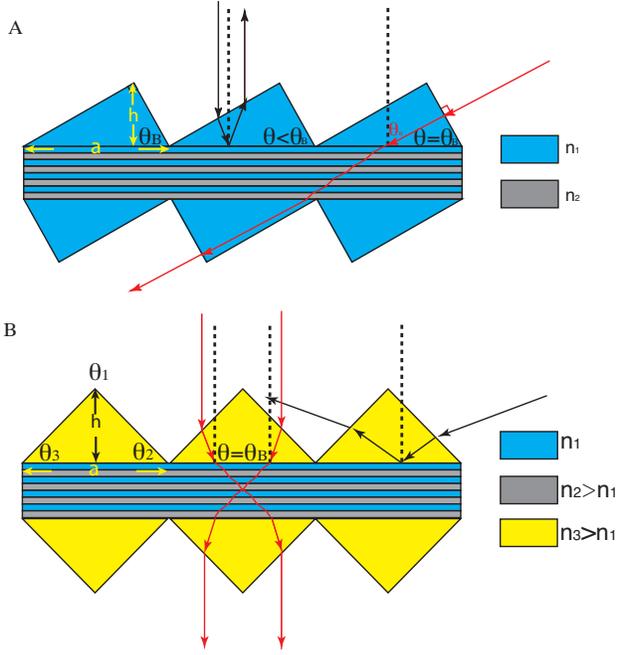}
\caption{\textbf{Illustration of optical angular selective filter in air using prisms} (A) Oblique incidence prism design (B) Normal incidence prism design. The red beams represent the light rays coming in at the right direction so the prisms can couple the light into the Brewster mode inside the PhC slab. The black beams represent the light coming in from other directions so they cannot be coupled into the Brewster mode inside the PhC slab; hence they are reflected back into the prisms.}
\label{fig:fig2}
\end{center}
\end{figure}

Here, we propose an alternative approach, using macroscopic prisms to couple propagating light in the air and the Brewster modes in the PhC. A simple prism design is illustrated in Fig.~1A, where the angular selective PhC filter demonstrated in \cite{Shen28032014} is sandwiched by two dielectric masks with periodic triangular patterns. One can choose the material that has the same refractive index as one of the layers in the PhC slab. The geometry of the prism is designed to eliminate the refraction from air into the low-index material at the transparent angle. Light incident at $\theta_{B,n_1}$ is normal to the prism/air interface, hence will pass through the whole system. On the other hand, light incident from other directions will undergo different refractions, and will be reflected by the PhC slab.

An experimental demonstration of the prism implementation (Fig.~1A) is shown in Fig.~2. The prism is fabricated with Acrylic ($n=1.44$), and the surface of the prism is mechanically and chemicaly polished to enhance optical quality. The periodicity of the triangular pattern is $a=2$ mm and $\theta_B=55^{\circ}$. The PhC filter is adopted from \cite{Shen28032014} and embedded in the prism; the two materials used to make the PhC filter are SiO$_2$ ($n=1.45$) and Ta$_2$O$_5$ ($n=2.08$). The overall transparency of the system is up to 68\% to $p$-polarized incident light at $\theta_B=55^{\circ}$ (Fig.~2C); the angular window of transparency is about 8$^{\circ}$. While the peak transmittance of the sandwiched PhC filter itself is 98\% \cite{Shen28032014}, the peak transmittance of the whole system is weakened by implementing the prism. The air-prism interfaces on two sides of the system cause a 7\% reflection, which may be eliminated by applying an anti-reflective coating on the interfaces. The remaining losses are due to the imperfection of the prism fabrication. Better cutting and polishing of the prism can reduce scattering loss. It reflects light at all other angles over the entire visible spectrum (Fig.~2B,D). The $p$-polarized transmittance of the sample in the visible spectrum was measured using an ultraviolet-visible spectrophotometer (Cary 500i) and presented in Fig.~2E; Although the sample no longer behaves like a mirror at off-Brewster angle due to image distortion of the prism, the transmission spectrum (and reflection spectrum) is very close to the spectrum of the sample immersed in index-matched liquid (  in \cite{Shen28032014}). 

\begin{figure*}[htbp]
\begin{center}
\includegraphics[width=5in]{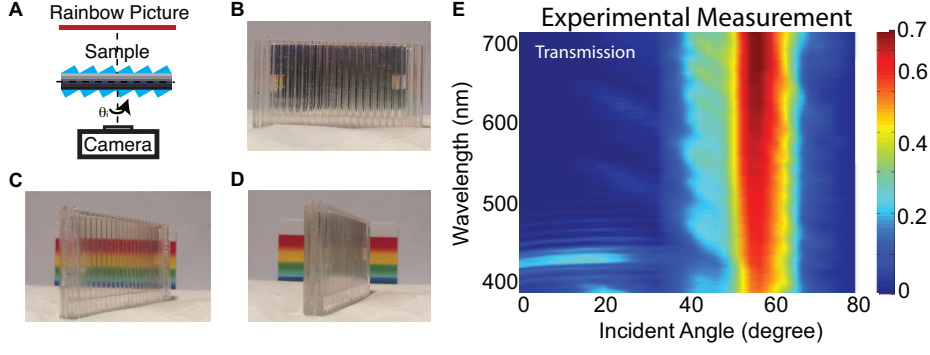}
\caption{\textbf{Demonstration of achieving angular selectivity in air} (A) Schematic view of the system setup. (B) Normal incident angle setup: the sample is reflective. (C) $\theta=55^{\circ}$ setup: the sample becomes transparent. (D) $\theta=75^{\circ}$ setup: the sample become reflective again. (E) Experimental measurement of the transmission spectrum of the prism-embedded sample.}
\label{fig:fig3}
\end{center}
\end{figure*}

One can change the angle of transparency of the system with careful design and alignment of the prisms on the top and bottom of the PhC slab. To show this, we propose the second design, which achieves angular selective window at normal incidence (Fig.~1(B)). In this design, the prisms are made up of materials having refractive index $n_3\approx2.1$ (e.g, L-BBH2 glass), and the three angles are $\theta_1=65^{\circ}$, $\theta_2=\theta_3=57.5^{\circ}$. With such a design, the light propagation direction inside the prism is always parallel to one edge of the prism; hence no light will be lost through the prism system. Detailed ray optics analysis that shows how this work is provided in the following section.

\section{Normal-incidence prism design}

One important thing to consider in designing the prism is that the prism itself should not block the light that is supposed to be transmitted at the transmission angle. Here we propose a prism design that uses L-BBH2 glass ($n_{L-BBH2}\approx2.1$ in the whole visible spectrum) to couple the light from normal incidence into the Brewster mode.

\begin{figure}[htbp]
\begin{center}
\includegraphics[width=3in]{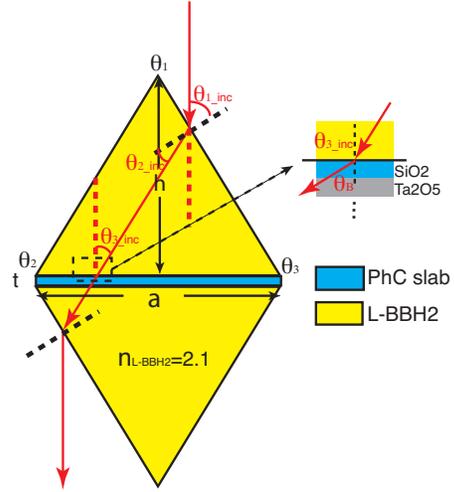}
\caption{\textbf{Normal-incidence prism design (one unit block)} The red lines represent the light path inside the prism/slab system of normally incident light.}
\label{fig:figureS1}
\end{center}
\end{figure}

As shown in Fig.~\ref{fig:figureS1}, the prism is composed of a periodic array of isosceles triangles with $\theta_1=65^{\circ}$ and $\theta_2=\theta_3=57.5^{\circ}$. In this case, from Snell's law, the refractive angle can be calculated by:

\begin{equation} 
n_{\mbox{\tiny{L-BBH2}}}\sin\theta_{2\_inc}=n_{air}\sin\theta_{1\_inc}
\end{equation}
where
\[\theta_{1\_inc}=90^{\circ}-(90^{\circ}-\theta_3)=\theta_3=57.5^{\circ}\]
therefore
\[\theta_{2\_inc}=\arcsin(\frac{n_{air}}{n_{L-BBH2}}\sin\theta_{1\_inc})=23.7^{\circ}.\]
Notice in this case the refracted ray is almost parallel to the left edge of the prism (it will be exactly parallel if $\theta_{2\_inc}=25^{\circ}$), hence most of the light coming in through the right edge of the prism will reach the Photonic Crystal (PhC) slab (instead of reaching the other edge of the prism). The same argument applies to the light coming in through the left edge of the prism.
Furthermore, from Fig.~\ref{fig:figureS1},
\[\theta_{3\_inc}=\theta_{1\_inc}-\theta_{2\_inc}=33.8^{\circ},\]
hence the angle of the ray coupled into the PhC slab (defined in SiO$_2$) can be calculated by:
\begin{equation} 
\theta_{SiO_2}=\arcsin(\frac{n_{L-BBH2}}{n_{SiO_2}}\sin\theta_{3\_inc})=53.7^{\circ}\approx\theta_{B,SiO_2}
\end{equation}

This shows the prism couples normal incident light into the Brewster angle of propagation inside the PhC slab. As mentioned in the main manuscript, light propagating at the Brewster angle inside the PhC slab will go through the whole PhC slab with almost no loss. 

Since the PhC slab has a thickness ($\approx5\mu m$) that is much smaller than the prism scale ($\approx1mm$), the PhC slab can be considered as a thin film angular selective filter in the prism systems, which allows light coming in at only the Brewster angle to pass while reflecting all other light. 

Finally, after the light ray has passed through the PhC slab, the light ray will undergo the same refraction and leave the lower prism at the same direction as it enters the upper prism (Fig.~\ref{fig:figureS1}), since the lower prism is identical to the upper prism. Notice that since the light propagation direction inside the prism is always parallel to one edge of the prism, all the light coming in through the left edge of the upper prism will reach the right edge of the bottom prism, while all the light coming in through the right edge of the upper prism will reach the left edge of the bottom prism. Hence no light will be lost through the prism system.

On the other hand, light coming in from other directions will undergo different refraction, and $\theta_{3\_inc}$ will no longer be $33.8^{\circ}$, thus will be reflected back by the angular selective PhC slab.

\section{Image preserved normal incident angular selective system design}

Although the prism designed in the above section successfully filters the light coming in at directions other than the normal incidence, it "flips" the light rays after they go through the prism system. As a result, the image before the prism will not be preserved after it goes through the prism. This is unacceptable for imaging applications.

\begin{figure}[htbp]
\begin{center}
\includegraphics[width=3.5in]{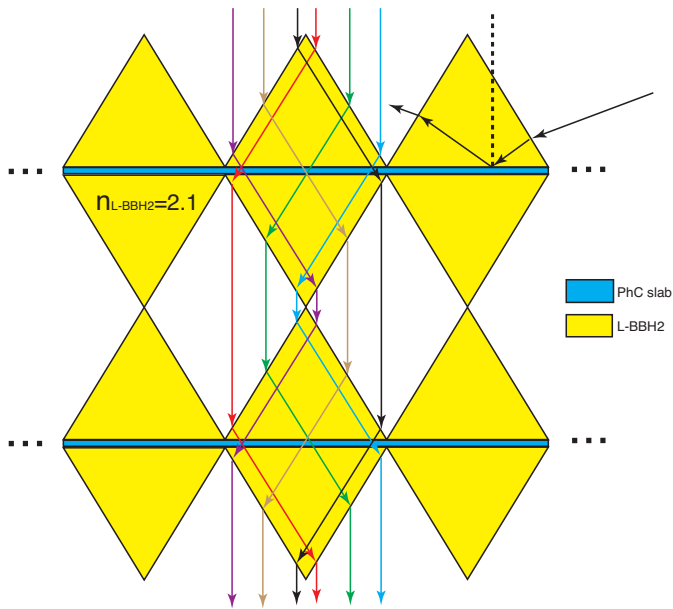}
\caption{\textbf{Normal-incidence, image-preserving angular selective system design.} Light rays are marked with different colors so that one can track where the incident light rays comes out after the whole system. At normal incidence, the transmitted light rays follow the same location as the incident light rays. Hence the image before and after the prism system will be preserved.}
\label{fig:figureS2}
\end{center}
\end{figure}

One can restore the image by placing another identical prism-PhC system next to the first one. As demonstrated in Fig.~\ref{fig:figureS2}, the "flipped" light ray will be "flipped" again after the second prism system. In this case, the transmission image will be fully restored.

However, up to now we have not thought of a prism design that preserves both the transmitted image and the reflected image.

\section{Conclusion}
Based on the key concept of optical broadband angular selectivity proposed in \cite{Shen28032014}, with the aid of a prism coupler, this report demonstrate the first experimental realization of an air-compatible optical broadband angular selective filter. We show that the angle of transparency is also adjustable and we provide a second design for normal incident angular selective filter.

\vspace{24pt}
\noindent {\bf Acknowledgments: }
This work was partially supported by the Army Research Office through the ISN under Contract Nos.~W911NF-13-D0001. The fabrication part of the effort, as well as (partially) M.S. were supported by the MIT S3TEC Energy Research Frontier Center of the Department of Energy under Grant No. DE-SC0001299.

\bibliography{Yichen_bib}

\end{document}